\documentclass[sigconf,natbib=true,anonymous=false]{acmart}

\usepackage{graphicx}
\usepackage{subcaption}
\usepackage{algorithm}
\usepackage{algpseudocode}
\usepackage{enumitem}
\usepackage{multirow}
\usepackage{tabularx}
\usepackage{threeparttable}
\usepackage{color, xspace}
\usepackage{appendix}

\AtBeginDocument{%
    \providecommand\BibTeX{{%
            \normalfont B\kern-0.5em{\scshape i\kern-0.25em b}\kern-0.8em\TeX}}}
\copyrightyear{2025}
\acmYear{2025}
\setcopyright{acmlicensed}\acmConference[SIGIR '25]{Proceedings of the 48th
International ACM SIGIR Conference on Research and Development in
Information Retrieval}{July 13--18, 2025}{Padua, Italy}
\acmBooktitle{Proceedings of the 48th International ACM SIGIR Conference on
Research and Development in Information Retrieval (SIGIR '25), July 13--18,
2025, Padua, Italy}
\acmDOI{10.1145/3726302.3730087}
\acmISBN{979-8-4007-1592-1/2025/07}

\usepackage{color, xspace}

\begin{document}

\title{STAR-Rec: Making Peace with Length Variance and Pattern Diversity in Sequential Recommendation}

\author{Maolin Wang}
\authornote{Both authors contributed equally to this research.}
\affiliation{%
  \institution{City University of Hong Kong}
  \streetaddress{}
  \city{}
  \state{}
  \country{}
}
\email{morin.wang@my.cityu.edu.hk}

\author{Sheng Zhang}
\authornotemark[1] 
\affiliation{%
  \institution{City University of Hong Kong}
  \streetaddress{}
  \city{}
  \state{}
  \country{}
}
\email{szhang844-c@my.cityu.edu.hk}

\author{Ruocheng Guo}
\affiliation{%
  \institution{Independent Researcher}
  \streetaddress{}
  \city{}
  \state{}
  \country{}
}
\email{rguo.asu@gmail.com}

\author{Wanyu Wang}
\authornote{Corresponding Author}
\affiliation{%
  \institution{City University of Hong Kong}
  \streetaddress{}
  \city{}
  \state{}
  \country{}
}
\email{wanyuwang4-c@my.cityu.edu.hk}

\author{Xuetao Wei}
\affiliation{%
  \institution{ Southern University of \\Science and Technology}
  \streetaddress{}
  \city{}
  \state{}
  \country{}
  \postcode{}
}
\email{weixt@sustech.edu.cn}

\author{Zitao Liu}
\affiliation{%
  \institution{Jinan University}
  \streetaddress{}
  \city{}
  \state{}
  \country{}
  \postcode{}
}
\email{zitao.jerry.liu@gmail.com}

\author{Hongzhi Yin}
\affiliation{%
  \institution{The University of Queensland}
  \streetaddress{}
  \city{}
  \state{}
  \country{}
  \postcode{}
}
\email{db.hongzhi@gmail.com}

\author{Yi Chang}
\affiliation{%
  \institution{Jilin University}
  \streetaddress{}
  \city{}
  \state{}
  \country{}
  \postcode{}
}
\email{yichang@jlu.edu.cn}

\author{Xiangyu Zhao}
\affiliation{%
  \institution{City University of Hong Kong}
  \streetaddress{}
  \city{}
  \state{}
  \country{}
  \postcode{}
}
\email{xianzhao@cityu.edu.hk}

\renewcommand{\shortauthors}{Wang et al.}

\begin{abstract}
Recent deep sequential recommendation models often struggle to effectively model key characteristics of user behaviors, particularly in handling sequence length variations and capturing diverse interaction patterns. We propose STAR-Rec, a novel architecture that synergistically combines preference-aware attention and state-space modeling through a sequence-level mixture-of-experts framework. STAR-Rec addresses these challenges by: (1) employing preference-aware attention to capture both inherently similar item relationships and diverse preferences, (2) utilizing state-space modeling to efficiently process variable-length sequences with linear complexity, and (3) incorporating a mixture-of-experts component that adaptively routes different behavioral patterns to specialized experts, handling both focused category-specific browsing and diverse category exploration patterns. We theoretically demonstrate how the state space model and attention mechanisms can be naturally unified in recommendation scenarios, where SSM captures temporal dynamics through state compression while attention models both similar and diverse item relationships. Extensive experiments on four real-world datasets demonstrate that STAR-Rec consistently outperforms state-of-the-art sequential recommendation methods, particularly in scenarios involving diverse user behaviors and varying sequence lengths. The implementation code is available anonymously online for easy reproducibility~\footnote{https://github.com/Applied-Machine-Learning-Lab/Star-Rec}.
\end{abstract}



\maketitle

\section{Introduction}

Sequential recommendation has evolved into a cornerstone technology across digital platforms~\cite{GRU4Rec,bert4rec,wang2024rethinking,DL4,liang2023mmmlp,wang2023multi,zhao2018deep}, transforming how users discover content and products on e-commerce sites, streaming platforms, and social networks~\cite{chen2023survey,gao2024smlp4rec,zhao2018recommendations,boka2024survey}. This field has witnessed paradigm shifts in modeling approaches, from initial recurrent architectures to self-attention mechanisms and recently to state-space formulations~\cite{gao2024smlp4rec,GRU4Rec,bert4rec,lightsan,mamba4rec}. A fundamental challenge lies in effectively processing users' complex interaction histories, where long-term preferences and short-term interests interweave~\cite{cui2024context,DL4,strec,bert4rec,GRU4Rec,Linrec}. While recent advances have improved temporal pattern capture~\cite{smlp4rec,mamba4rec,liu2024bidirectional}, comprehensively modeling the dynamic nature of user behaviors while preserving crucial historical signals remains an open challenge~\cite{smlp4rec,mlp4rec,mamba4rec}.

Specifically, we observe that in real-world sequential recommendation scenarios, user behaviors naturally exhibit three key characteristics: (1) \textbf{Items Pattern Diversity} shows that users tend to interact with items that share inherent similarities (\emph{e.g.}, browsing different styles of sneakers) while also exploring items across various categories (\emph{e.g.}, switching between electronics, clothing, and books), demonstrating complex item-level interaction patterns~\cite{zhao2023embedding,wen2019building,zhang2024dns}. (2) \textbf{Length Variance} reflects that user interaction sequences demonstrate significant variation in length, ranging from new users with only a few interactions to active users with hundreds of historical behaviors, which poses challenges for unified sequence modeling~\cite{liu2024bidirectional,liu2024sequential,qu2024survey}. (3) \textbf{Behavioral Pattern Diversity} indicates that sequences often contain diverse behavioral patterns that reflect changing interests and intentions over time; users may alternate between focused category-specific browsing (\emph{e.g.}, intensively searching for sneakers) and diverse category exploration (\emph{e.g.}, browsing across different product categories) or switch between casual browsing and purposeful purchasing behaviors \cite{zhang2022hierarchical,chen2023survey}.

Based on these observations, sequential recommender systems need to address three fundamental challenges.
First, the diverse item patterns require effective modeling of inherent item relationships. These relationships provide essential signals for recommendation but are often overlooked by pure sequential models focusing solely on temporal patterns~\cite{DL4,Embedding}.
Second, the significant variance in sequence lengths poses a major challenge. While state space models (SSMs) show impressive efficiency in handling long sequences~\cite{mamba4rec,Linrec}, they struggle with short sequences due to unstable state estimation, leading to poor recommendations for new users or users with sparse interactions~\cite{gu2023mamba,MambaRec}.
Third, the diverse behavioral patterns reflecting multiple user interests present a modeling challenge. Existing methods (RNNs, Transformers, and SSMs) process sequences uniformly, failing to adapt to varying patterns: RNNs~\cite{GRU,GRU2,GRU3} focus on immediate dependencies, Transformers~\cite{vaswani01,lightsan,sse-pt,wang2024large} treat all positions equally, and SSMs maintain a single state representation - none explicitly models multiple behavior patterns~\cite{GRU4Rec,bert4rec,mamba4rec}.
These limitations of current approaches are particularly evident: RNN-based methods show suboptimal performance~\cite{GRU4Rec,narm}, Transformers face quadratic complexity issues~\cite{vaswani01,Linformer}, and SSMs~\cite{qu2024survey,gu2023mamba}, while efficient, perform poorly on short sequences and lack mechanisms for diverse patterns~\cite{MambaRec,mamba4rec,liu2024bidirectional}. This necessitates a solution that addresses sequence length variation and pattern diversity within complex user-item interactions while maintaining efficiency.

To address these challenges, we propose STAR-Rec~(\textbf{ST}ate-space and preference-aware-\textbf{A}ttention based \textbf{R}ecommendation), a novel architecture that synergistically combines preference-aware attention and state-space modeling through a sequence-level mixture-of-experts (MoE) framework. The preference-aware-attention mechanism captures static item relationships and preference patterns, complementing the state-space model, which efficiently processes temporal dynamics with linear complexity. The MoE component adaptively routes different behavioral patterns to specialized experts, enabling more nuanced modeling of user behaviors. This unified framework effectively leverages both similarity-based and temporal information while maintaining computational efficiency through selective computation paths.

\begin{figure*}[t]
    \centering
    \includegraphics[width=0.9\linewidth]{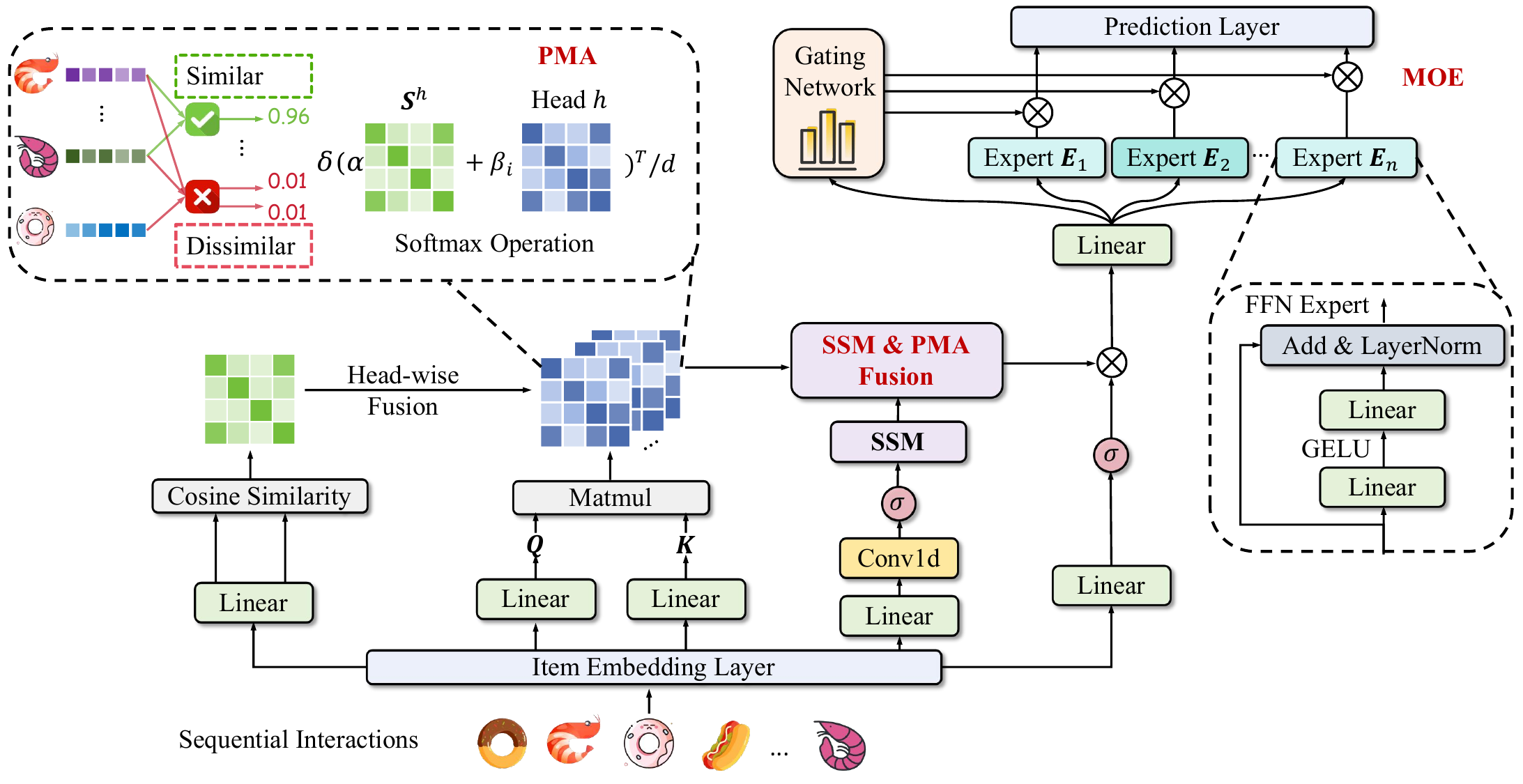}
    \vspace{-10pt}
    \caption{Overview of STAR-Rec architecture. The model consists of (1) an item embedding layer for initial representations, (2) a unified modeling block that combines preference-aware multi-head attention (PMA) for preference aggregation and state-space model (SSM) for temporal dynamics through an adaptive fusion mechanism, and (3) a mixture-of-experts layer for specialized pattern recognition and prediction.}
    \label{fig:overview}
\end{figure*}
Our main contributions are summarized as follows:

\begin{itemize}[leftmargin=*]

\item We propose STAR-Rec, a novel sequential recommendation framework that synergistically integrates preference-aware multi-head attention and selective state-space modeling with mixture-of-experts for capturing complex user behavior patterns.

\item We theoretically demonstrate that SSM and attention mechanisms can be naturally unified in recommendation scenarios, where SSM captures temporal dynamics through state compression while preference-aware multi-head attention captures static item relationships. This complementary nature enables their effective fusion through adaptive weighted combinations.

\item We conduct comprehensive and extensive experiments on four real-world datasets to evaluate STAR-Rec, and experimental results demonstrate its superior performance over state-of-the-art sequential recommendation methods.
\end{itemize}

\section{Preliminaries}
In this section, we first introduce the sequential recommendation task and then present the fundamental concepts of State Space Models (SSMs), which serve as key components in our framework.
\subsection{Sequential Recommendation}
Sequential recommendation aims to predict users' next interactions based on their historical behavior sequences~\cite{DL4,GRU4Rec}. This task is fundamental in various applications such as e-commerce and content platforms, where understanding and predicting user preferences can significantly enhance personalization quality~\cite{bert4rec,strec}. Given a set of users $\mathcal{U} = \{u_1, u_2, \dots, u_{|\mathcal{U}|}\}$ and items $\mathcal{V} = \{v_1, v_2, \dots, v_{|\mathcal{V}|}\}$, each user $u_i$ has an interaction sequence $s_i = [v_1^{(i)}, v_2^{(i)},\dots, v_{n_i}^{(i)}]$ of length $n_i$. The goal is to predict the next item the user will interact with based on their historical sequence~\cite{Kang01,Linrec}. This task presents unique challenges in modeling both temporal dependencies and item relationships, requiring a balanced approach to capture both sequential patterns and item similarities~\cite{Embedding,smlp4rec}.
\subsection{State Space Models (SSMs)}

State Space Models (SSMs)~\cite{dao2024transformers,HiPPOs21,mamba4rec,gu2023mamba,qu2024survey} provide a powerful mathematical framework for modeling temporal dynamics through linear ODEs~\cite{Hamilton20}. Recently, efficient SSM variants like Mamba have shown promising results in sequence modeling tasks. In continuous time, SSMs can described as follows:
$$
\boldsymbol{h}'(t) = \boldsymbol{P} \boldsymbol{h}(t) + \boldsymbol{Q} \boldsymbol{x}(t), \quad
\boldsymbol{y}(t) = \boldsymbol{C} \boldsymbol{h}(t),
$$
where $\boldsymbol{P} \in \mathbb{R}^{N \times N}$, $\boldsymbol{Q} \in \mathbb{R}^{N \times D}$, and $\boldsymbol{C} \in \mathbb{R}^{D \times N}$ are learnable matrices that control state transitions and output generation. For practical implementation with discrete sequences, the model computation mechanism becomes:
$$
\boldsymbol{h}_{t} = \boldsymbol{\bar{P}} \boldsymbol{h}_{t-1} + \boldsymbol{\bar{Q}} \boldsymbol{x}_t, \quad
\boldsymbol{y}_t = \boldsymbol{C} \boldsymbol{h}_t,
$$
where $\boldsymbol{\bar{P}} = \exp(\Delta \boldsymbol{P})$ and $\boldsymbol{\bar{Q}} = (\Delta \boldsymbol{P})^{-1}(\exp(\Delta \boldsymbol{P}) - \boldsymbol{I})\Delta \boldsymbol{Q}$. T
The State-Space Model~(SSM) compresses the variable-length input sequence $\{\boldsymbol{x}(\tau) \mid \tau \leq t\}$ into a high-capacity, information-dense state $\boldsymbol{h}_t$ through mappings defined by matrices $\bar{\boldsymbol{P}}$ and $\bar{\boldsymbol{Q}}$, where $N$ denotes the state-space hidden dimension.

The output $\boldsymbol{y}_t$ is then generated via a mapping $\phi(\boldsymbol{h}_t \mid \boldsymbol{C}) \to \boldsymbol{y}_t$:
$$
f\big(\{\boldsymbol{x}(\tau) \mid \tau \leq t\} \;\big|\; \bar{\boldsymbol{P}}, \bar{\boldsymbol{Q}}\big) \;\rightarrow\; \boldsymbol{h}_t, \quad 
\phi(\boldsymbol{h}_t \mid \boldsymbol{C}) \;\rightarrow\; \boldsymbol{y}_t.
$$
The model's behavior over multiple time steps creates an implicit attention mechanism where past inputs influence current outputs through state transitions. This property makes SSMs particularly suitable for modeling sequential dependencies~\cite{qu2024survey}.

\section{Methodology}
In this section, we present the STAR-Rec framework, including the model architecture and its key components.

\subsection{Model Architecture}

The proposed model, \textbf{STAR-Rec}, integrates state-space modeling with attention mechanisms to unify their strengths, offering a robust solution for the sequential recommendation. This architecture is designed to efficiently handle long-range dependencies while incorporating item similarities, which are often overlooked in traditional approaches. Figure~\ref{fig:overview} illustrates the overall design of STAR-Rec.

\subsubsection{\textbf{Item Embedding Layer}}

The item embedding layer is a foundational component that transforms discrete items into dense continuous representations, enabling the model to capture intricate item relationships. Given an input sequence $s = [v_1, v_2, \dots, v_T]$ of length $T$, each item is embedded into a $d$-dimensional space:
$$
\boldsymbol{L}^{(embed)} = \text{ItemEmbedding}(s) \in \mathbb{R}^{T \times d}.
$$

For long-term sequential recommendation, information about both items and their positions should be encoded into the model. We denote the length of input user-item interactions as $N$ and embedding size as $d$. For a user $u_i$ who has an interaction sequence $s_i=[v_1, v_2, \dots, v_t, \dots v_{n_i}]$, the $t$-th item $v_t\in \mathbb{R}^{D_t}$ and its position $p_t\in \mathbb{R}^{D_t}$ can be projected into a dense representation $\boldsymbol{e}_t^s$ and, $\boldsymbol{e}_t^p$ respectively, through an embedding layer:
$$
\boldsymbol{e}_t^s = \boldsymbol{W}_t^s v_t, \ \ \boldsymbol{e}_t^p = \boldsymbol{W}_t^p p_t,
$$
where $\boldsymbol{W}_t^s \in \mathbb{R}^{d\times D_t}, \boldsymbol{W}_t^p \in \mathbb{R}^{d\times D_t}$ are trainable weighted matrices for $t$-th item and positional embedding, and $D_t$ is its corresponding dimension. Finall,y the user-item interaction can be represented as:
$$
\boldsymbol{E}^{(embed)} = [\boldsymbol{e}_1^s+\boldsymbol{e}_1^p, \boldsymbol{e}_2^s+\boldsymbol{e}_2^p, \cdots, \boldsymbol{e}_N^s+\boldsymbol{e}_N^p]^{\mathrm{T}}.
$$

\subsubsection{\textbf{Preference-aware Multi-head Attention Path}}
In real-world recommendation scenarios, capturing dynamic user preferences is crucial yet challenging, as they need to be modeled alongside static item similarities~\cite{wen2019building,chen2023survey}. Traditional attention mechanisms typically model global item relationships without considering user preferences~\cite{liu2024bidirectional,vaswani01}. Such approaches treat all historical interactions uniformly, failing to capture the varying importance of different items in user preference representation.

We propose a Preference-aware Multi-head Attention (PMA) mechanism that distinctly models both preference dynamics and item similarities. The preference awareness is achieved through a transformation that explicitly captures user preference patterns, where each raw item representation $\boldsymbol{X}_{\text{raw}} \in \mathbb{R}^{L \times D}$ (where $L$ denotes the maximum padded sequence length and $D$ is the feature dimension) undergoes a head-specific gating:
$$
\boldsymbol{X}_m^h = \boldsymbol{X}_{\text{raw}} \odot \sigma(\boldsymbol{X}_{\text{raw}}\boldsymbol{W}_g^h + \boldsymbol{b}_g^h),
$$
where $\boldsymbol{W}_g^h \in \mathbb{R}^{D \times D}$ and $\boldsymbol{b}_g^h \in \mathbb{R}^D$ learn to identify preference-relevant features for each attention head $h$. This transformation enables each attention head to focus on different aspects of user preferences.

To balance between user preference dynamics and inherent item relationships that cannot be effectively encoded in state space $h(t)\in \mathbb{R}^{L \times D}$, we compute dedicated similarity matrices for each head as the following equation:
$$
\boldsymbol{S}^h \in \mathbb{R}^{L \times L} = \text{CosSim}(\boldsymbol{X}_{raw}, \boldsymbol{X}_m^h).
$$

These relationships are further refined through a preference-aware thresholding mechanism per head:
$$
\boldsymbol{S}_{ij}^h = \begin{cases}
\boldsymbol{S}_{ij}^h, & \text{if } \boldsymbol{S}_{ij}^h \geq \tau^h, \\ 0.01, & \text{otherwise}.
\end{cases}
$$

The core of our preference-aware design lies in the dual-path attention computation that separately models preference dynamics and item similarities. 
For each head $h$, the attention scores integrate both components and can be expressed as:
$$
\boldsymbol{A}_h^{attn} \in \mathbb{R}^{L \times L} = \text{Softmax}\left(\frac{w_1^h\boldsymbol{Q}_h^{attn}\boldsymbol{K}_h^T + w_2^h\boldsymbol{S}^h}{\sqrt{D_k}}\right),
$$
where $\boldsymbol{Q}_h^{attn}\boldsymbol{K}_h^T \in \mathbb{R}^{L \times L}$ captures temporal preference evolution through query-key interactions while $\boldsymbol{S}^h \in \mathbb{R}^{L \times L}$ maintaining complementary item relationships. The learnable weights $w_1^h$ and $w_2^h$ enable adaptive balancing between dynamic and static components, with $\boldsymbol{Q}_h^{attn}, \boldsymbol{K}_h \in \mathbb{R}^{L \times D_k}$, and $D_k = D/H$ being the dimension of keys per head. These weights are adaptively learned via:
$$
w_i^{h,(t)} = \mathrm{softmax}(w_i^{h,(t-1)}) = \frac{e^{w_i^{h,(t-1)}}}{e^{w_1^{h,(t-1)}}+e^{w_2^{h,(t-1)}}}, \ \ i=1,2,
$$

The output of each attention head is computed as:
$$
\boldsymbol{Y}_{attn}^h \in \mathbb{R}^{L \times D_k} = \boldsymbol{A}_h^{attn}\boldsymbol{V}_h,
$$
where $\boldsymbol{V}_h \in \mathbb{R}^{L \times D_k}$.
The value matrix $\boldsymbol{V}_h$ is obtained by projecting input $\boldsymbol{X}$ through a learnable linear transformation $\boldsymbol{W}_v^h$:
$$
\boldsymbol{V}_h = \boldsymbol{X}\boldsymbol{W}_v^h, \ \ \boldsymbol{W}_v^h \in \mathbb{R}^{D \times D_k}.
$$

The final multi-head attention output combines the preference-aware representations from all heads through a linear projection:
\begin{align}\notag 
\boldsymbol{Y}_{\text{attn}} &= \text{Concat}(\boldsymbol{A}_1^{\text{attn}}\boldsymbol{X}\boldsymbol{W}_v^1, \boldsymbol{A}_2^{\text{attn}}\boldsymbol{X}\boldsymbol{W}_v^2, \ldots, \boldsymbol{A}_H^{\text{attn}}\boldsymbol{X}\boldsymbol{W}_v^H) \\ \label{eq:1}
&= \text{Concat}(\boldsymbol{Y}_{\text{attn}}^1, \boldsymbol{Y}_{\text{attn}}^2, \ldots, \boldsymbol{Y}_{\text{attn}}^H) \cdot \boldsymbol{W}^O,
\end{align}
where $\boldsymbol{W}^O \in \mathbb{R}^{(H \times D_k) \times D}$ is the output projection matrix.
Our design enables the model to capture diverse preference patterns through two complementary mechanisms. The attention mechanism learns global item relationship patterns from the entire training dataset, discovering various types of item associations like category-based, functionality-based, or popularity-based patterns. Meanwhile, the similarity component captures ~\textbf{items pattern diversity} by modeling user-specific patterns, such as individual browsing habits, purchase frequencies, and temporal preferences, effectively serving as a personalized user profile. 

\begin{figure}[t]
    \centering
    \includegraphics[width=0.95\linewidth]{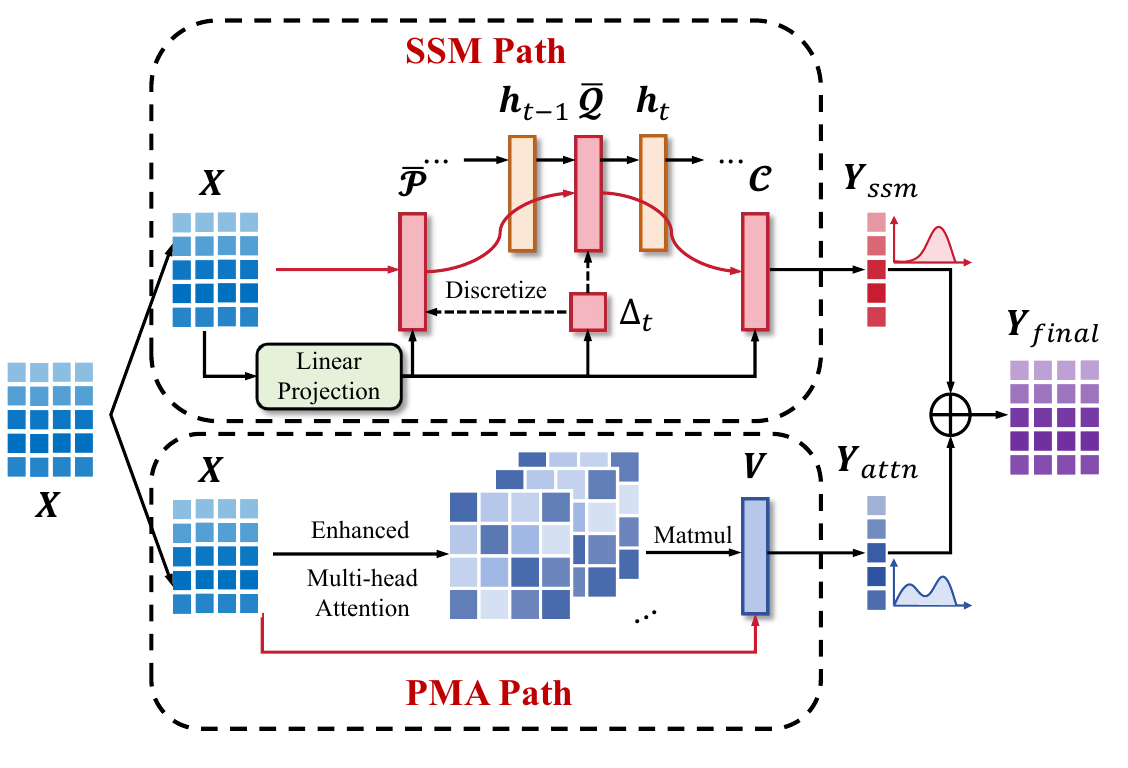}
    \caption{Overview of the adaptive fusion mechanism integrating SSM and PMA paths for sequence modeling.}
    \label{fig:fusion1}
    \vspace{-10pt}
\end{figure}
\subsubsection{\textbf{Adaptive Fusion of SSM and PMA}}

\subsubsection{\textbf{State-space Model Path}}

Sequential recommendation faces unique challenges in modeling long-range dependencies with varying temporal dynamics~\cite{zhao2023embedding,wen2019building,liu2024bidirectional,qu2024survey,chen2023survey}, where traditional attention mechanisms suffer from quadratic complexity~\cite{Keles7On} and standard State Space Models (SSMs) struggle to retain relevant preference information~\cite{qu2024survey}. We adopt SSMs as one modeling path in our framework, leveraging their linear computational complexity and strong temporal pattern modeling capabilities to serve as an efficient backbone for sequential recommendation scenarios~\cite{mamba4rec,qu2024survey}. 

For such scenarios, we extend the formulation to tensor form. Here, we can derive the output $\boldsymbol{Y}_{\text{ssm}} \in \mathbb{R}^{L \times D}$:
$$
    \boldsymbol{Y}_{\text{ssm}}^{(t,d)} = \sum_{k=0}^t \boldsymbol{C}_{t,:} \bar{\mathcal{P}}_{t,d,:,:} \cdots \bar{\mathcal{P}}_{k,d,:,:} \bar{\mathcal{Q}}_{t,:,d}\boldsymbol{X}^{(k,d)},
$$
where $\mathcal{C} \in \mathbb{R}^{L \times N}$, $\bar{\mathcal{Q}} \in \mathbb{R}^{L \times N \times D }$, and $\boldsymbol{X}^{(k,d)} \in \mathbb{R}$ represents the $k$-th input embedding, $t\in [L]$. The $\bar{\mathcal{P}} \in \mathbb{R}^{L \times D \times N \times N}$ matrix plays a crucial role in enhancing the state. Through successive multiplications, $\bar{\mathcal{P}}_{:,d,:,:}$ acts as a mapper that continuously compresses and aggregates historical information.

To better understand the structure, we define a tensor $\mathcal{M} \in \mathbb{R}^{D \times L \times L}$ can be expressed as follows:
$$
\mathcal{M}_{d,t,k} = \boldsymbol{C}_{t,:} \bar{\mathcal{P}}_{t,d,:,:} \cdots \bar{\mathcal{P}}_{k,d,:,:} \bar{\mathcal{Q}}_{t,:,d}.
$$
Therefore, we can express $\boldsymbol{Y}_{\text{ssm}}$ as a concatenation operation:
\begin{align}\notag 
\boldsymbol{Y}_{\text{ssm}} &= \text{Concat}(\mathcal{M}_{1,:,:}\boldsymbol{X}\boldsymbol{L}^{(1)}, \mathcal{M}_{2,:,:}\boldsymbol{X}\boldsymbol{L}^{(2)}, \ldots, \mathcal{M}_{D,:,:}\boldsymbol{X}\boldsymbol{L}^{(D)}) \\ \label{eq:2} 
&= \text{Concat}(\hat{\boldsymbol{Y}}_{\text{ssm}}^{(:,1)}, \hat{\boldsymbol{Y}}_{\text{ssm}}^{(:,2)}, \ldots,  \hat{\boldsymbol{Y}}_{\text{ssm}}^{(:,D)}) \cdot \boldsymbol{I},
\end{align}
where $\hat{\boldsymbol{Y}}_{\text{ssm}}^{(:,d)} = \mathcal{M}_{d,:,:}\boldsymbol{X}\boldsymbol{L}^{(d)}$, $\boldsymbol{L}^{(d)} \in \mathbb{R}^{D \times 1}$ is a selection matrix, and $\boldsymbol{I} \in \mathbb{R}^{D \times D}$ is the identity matrix. $\mathcal{M}$ contains $D$ attention-like masks, with the number of heads $H$ being equal to the feature dimension $D$. Each slice $\mathcal{M}_{d,:,:}$ represents an independent attention pattern for the $d$-th feature dimension. SSM can thus be viewed as an attention mechanism where each feature dimension has its own attention pattern. However, relying solely on the $\bar{\mathcal{P}}$ and $\bar{\mathcal{Q}}$ for time-based compression faces challenges with mixed-length sequences. While long sequences benefit from SSM's temporal modeling, short sequences or rapidly changing preferences often lack clear temporal patterns and are driven more by short-term associations. To address this limitation, we enhance the SSM signal by incorporating preference values from the attention mechanism, allowing each channel $\boldsymbol{Y}_{ssm}$ to be represented in an attention form for better compression of dynamic patterns.

Sequential recommendation requires modeling both temporal dynamics and preference patterns effectively. While SSMs excel at processing long-range dependencies with linear complexity, they may miss essential preference signals in short sequences. This motivates us to design a unified framework that leverages SSM's efficiency and strength in modeling item relationships about attention.

As shown in Figure~\ref{fig:fusion1}, we introduce an adaptive fusion mechanism integrating SSM with preference-aware attention. This attention mechanism complements the SSM by capturing static preference relationships and item associations that temporal compression alone cannot model. By integrating cosine similarity into the attention computation, preference-aware attention assigns higher weights to more representative information in scenarios involving short-term behaviors or clusters of similar items.

The fused output combining SSM and PMA paths can be expressed as the following equation:
\begin{align}\notag
&\boldsymbol{Y}_{\text{final}}
= \gamma_1 \boldsymbol{Y}_{\text{ssm}} + \gamma_2 \boldsymbol{Y}_{\text{attn}} 
\\\notag&=
 \Bigl[
\text{Concat}\underbrace{\bigl(
\hat{\boldsymbol{Y}}_{\text{ssm}}^{(:,1)}, 
\hat{\boldsymbol{Y}}_{\text{ssm}}^{(:,2)}, 
\ldots,  
\hat{\boldsymbol{Y}}_{\text{ssm}}^{(:,D)}}_{\textit{Long-range heads~(SSM)}},
\underbrace{
\boldsymbol{Y}_{\text{attn}}^1,
\boldsymbol{Y}_{\text{attn}}^2,
\ldots,
\boldsymbol{Y}_{\text{attn}}^H}_{\textit{Preference Heads~(PMA)}}
\bigr)
\Bigr.\cdot
\\& \label{eq:3}
\Bigl.\text{diag}(\gamma_1 \boldsymbol{I};\gamma_2\boldsymbol{W}^O)
\Bigr]
\end{align}
where $\gamma_1$ and $\gamma_2$ are adaptively learned weights through softmax normalization. Following the concatenation operations in Eq.~\eqref{eq:1} and Eq.~\eqref{eq:2}, this unified mapping consolidates both modeling the two paradigms into a single framework.

\begin{figure}[t]
    \centering
    \includegraphics[width=\linewidth]{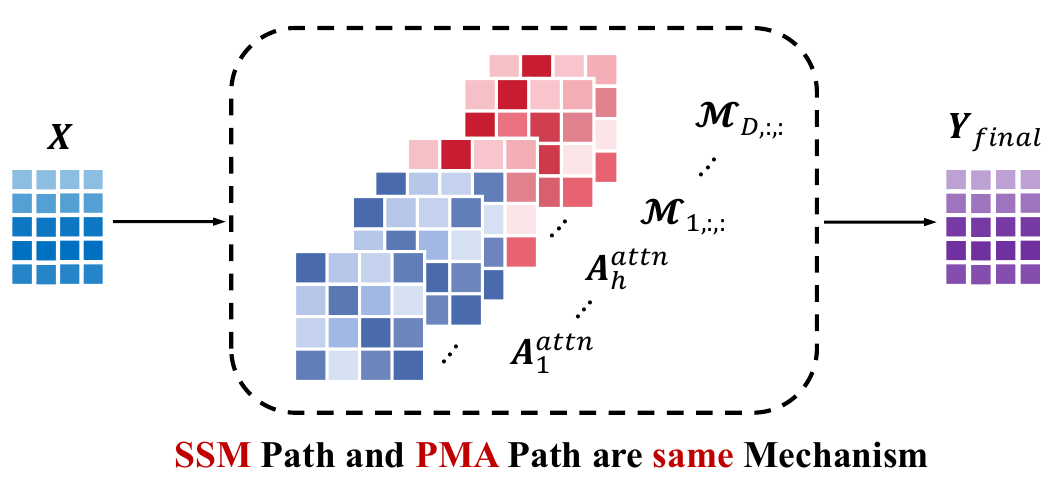}
    \vspace{-10pt}
    \caption{Illustration of the unified computational pattern between SSM and PMA paths, where both mechanisms can be represented as different attention heads through matrix multiplication operations, with SSM focusing on temporal dependencies and PMA emphasizing preference similarities.
    }
    \label{fig:fusion2}
        \vspace{-10pt}
\end{figure}

As illustrated in Figure~\ref{fig:fusion2} and Eq.~\eqref{eq:3}, both SSM and PMA architectures share fundamental computational patterns in sequence modeling. Given that SSMs can be formulated as implicit attention mechanisms through their state-transition dynamics~\cite{dao2024transformers}, and based on their unified forms in Eq.~\eqref{eq:1}, Eq.~\eqref{eq:2}, and Eq.~\eqref{eq:3}, both paths follow the same computational mechanism of concatenating multiple attention-like operations. The head-specific adaptive weighting mechanisms balance these diverse global and personalized patterns appropriately, allowing the model to capture rich, multi-faceted user preferences across different attention perspectives. This dual-path adaptive fusion effectively addresses both \textbf{item pattern diversity} and \textbf{length variance}:
SSM excels at processing long dependency patterns, while PMA specializes in capturing user-oriented static interaction patterns through direct attention mechanisms, enabling effective modeling even with extremely short sequences.

\subsubsection{\textbf{Mixture-of-Experts Prediction}}

The mixture-of-experts architecture addresses the challenge of modeling diverse behavioral patterns within user sequences. Each expert specializes in recognizing specific types of patterns through a distinct feed-forward neural network structure:
$$
\boldsymbol{E}_i(\boldsymbol{Y}) = \text{Linear}_2(\text{GELU}(\text{Linear}_1(\boldsymbol{Y}))),
$$
where different experts can develop specialized recognition capabilities for various sequential patterns, such as fine-grained preference-based preferences or complex temporal dynamics.

A gating network dynamically routes sequence representations to the most appropriate experts based on the input characteristics:
$$
\boldsymbol{G}_i = \text{Softmax}(\text{Linear}_g(\boldsymbol{Y}))_i,
$$
enabling adaptive pattern recognition for heterogeneous user behaviors (e.g., switching between casual browsing and purposeful purchasing).
The final prediction combines the specialized expert predictions as follows:
$$
\hat{\boldsymbol{Y}} = \sum_{i=1}^n G_i\boldsymbol{E}_i(\boldsymbol{Y}).
$$
This mixture-of-experts design directly addresses \textbf{behavioral pattern diversity} within individual sequences. Unlike traditional approaches that process sequences uniformly, the gating mechanism allows STAR-Rec to adaptively route different patterns to specialized experts. For instance, when a user exhibits preference-driven behaviors (e.g., browsing similar items), the relevant experts are activated. In contrast, temporal pattern experts are prioritized when sequential dependencies are more important.

\subsection{Training Objective}

The training of STAR-Rec is guided by a composite loss function that combines prediction accuracy with expert diversity. The overall objective can be expressed as:
$$
\mathcal{L} = \mathcal{L}_{\text{pred}} + \lambda\mathcal{L}_{\text{div}},
$$
where $\mathcal{L}_{\text{pred}}$ is the prediction loss, and $\mathcal{L}_{\text{div}}$ encourages diversity among experts, with $\lambda$ controlling their trade-off.

The prediction loss adopts binary cross entropy for the binary classification task. For implicit feedback, it predicts whether a user will interact with an item using the following formulation:
$$
\mathcal{L}_{\text{pred}} = -\sum_{i} [y^{(i)} \log {y}^{(i)} + (1-y^{(i)}) \log(1-{y}^{(i)})],
$$
where $y^{(i)} \in \{0,1\}$ represents the ground truth label indicating whether the user interacted with the item, and $\hat{y}^{(i)}$ is the predicted probability of positive interaction. The model learns to optimize this binary classification objective to distinguish between positive and negative interactions.

To ensure diversity among experts, a regularization term penalizes redundant gating weights:
$$
\mathcal{L}_{\text{div}} = \left\|\frac{\boldsymbol{G}^T\boldsymbol{G}}{B} - \frac{\boldsymbol{I}_n}{n}\right\|_F^2,
$$
where $\boldsymbol{G}$ is the gating weight matrix $B$ is the batch size, and $n$ is the number of experts. This promotes specialization and reduces redundancy, enhancing the model's ability to generalize across diverse user behaviors.

\subsection{Model Properties}

STAR-Rec addresses sequential recommendation challenges through three key properties:
(1) \textbf{Pattern-aware Integration}: Our model integrates static item similarities into dynamic preference modeling through PMA's dual-attention design, addressing the challenge of capturing both static and dynamic preference patterns through similarity-enhanced attention computation.
(2) \textbf{Adaptive Sequence Learning}: The framework combines SSM's efficient modeling of temporal dependencies with PMA's strength in capturing preference patterns, achieving robust performance across varying sequence lengths through adaptive fusion.
(3) \textbf{Behavioral Specialization}: The mixture-of-experts architecture enables specialized modeling of distinct interaction patterns through diverse expert networks, with a gating mechanism dynamically routing sequences to the most suitable experts based on behavioral characteristics, allowing adaptive pattern recognition for heterogeneous user behavior modeling.

\section{Experiments}
To validate the performance of our STAR-Rec framework, we designed and implemented a series of experimental studies. In this section, we first describe the experimental setup, followed by an in-depth analysis of the results. Specifically, our experiments aim to investigate the following research questions:
\begin{itemize}[leftmargin=*]
    \item \textbf{RQ1:} How does STAR-REC perform compared to state-of-the-art recommendation models on different datasets?

    \item \textbf{RQ2:} How does STAR-Rec perform when dealing with limited historical interaction information?
    \item \textbf{RQ3:} How do different components (PMA and MoE layers) contribute to STAR-REC's effectiveness?
    \item \textbf{RQ4:} How do the key hyperparameters affect the model's performance and efficiency?
    \item \textbf{RQ5:} How does our model compare to other baselines in terms of training and inference efficiency?
\end{itemize}

\subsection{Datasets and Evaluation Metrics}

As shown in Table~\ref{tab:data1}, we evaluate our model on four widely-used public datasets: MovieLens 1M~\footnote{https://grouplens.org/datasets/movielens/}, Amazon Beauty, Amazon Video Games, and Amazon Baby~\footnote{https://cseweb.ucsd.edu/~jmcauley/datasets.html\#amazon\_reviews}. MovieLens-1M is a stable benchmark dataset containing 1 million ratings from 6,000 users on 4,000 movies, with each user having at least 20 ratings. Each record includes user ID, movie ID, rating (1-5), and timestamp. We also utilize three Amazon category datasets: Beauty, Video Games, and Baby, which contain user-product interactions such as browsing, purchases, and ratings, along with product metadata. For each user's interaction sequence, we use the penultimate and last items for validation and testing, respectively, resulting in a 1:1 ratio between validation and test sets.
We employ three widely used metrics to evaluate recommendation performance: Recall@K, Mean Reciprocal Rank (MRR@K), and Normalized Discounted Cumulative Gain (NDCG@K). For data preparation, we first sort all user interactions chronologically and adopt the leave-one-out strategy~\cite{Linrec} for dataset splitting. Specifically, for each user's interaction sequence, we select the penultimate interaction for validation, making the validation set size equal to the number of users in our datasets. The last interaction is used for testing, while all previous interactions form the training set. 

\begin{table}[t]
    \caption{\textbf{Statistical Information of Adopted Datasets.}}
    \label{tab:data1}
    \renewcommand{\arraystretch}{1}
    \resizebox{\linewidth}{!}{
    \begin{tabular}{cccccc}
        \toprule
        Datasets& \# Users & \# Items &\# Interactions & Avg.Length & Sparsity\\
        \midrule
        ML-1M & 6,041 & 3,707 & 1,000,209 & 165.60 & 95.53\%\\
        Beauty & 22,364 & 12,102 & 198,502 & 8.88 & 99.93\%\\
        Baby  & 19,445  & 7,050 & 160,792 & 8.27 & 99.88\%\\
        Video Games& 24,304 & 10,673 & 231,780 & 9.54 & 99.91\%\\
        \bottomrule
    \end{tabular}}
    \vspace{-4mm}
    \end{table}

\begin{table*}[t]
    \caption{Overall performance comparison of sequential recommendation methods. The best results are in \textbf{bold} with marker $^\ast$ for statistical significance ($p<0.05$).}
    \label{tab:performance}
    \renewcommand{\arraystretch}{1.05}
    \resizebox{\linewidth}{!}{
    \begin{tabular}{ccccccccccccc}
        \toprule
        \multirow{2}{*}{Models} 
        & \multicolumn{3}{c}{ML-1M} 
        & \multicolumn{3}{c}{Amazon Beauty} 
        & \multicolumn{3}{c}{Amazon Baby} 
        & \multicolumn{3}{c}{Amazon Video Games}\\
        \cline{2-13}
        \vspace{-3mm}\\
        & Recall@10 & MRR@10 & NDCG@10 
        & Recall@10 & MRR@10 & NDCG@10 
        & Recall@10 & MRR@10 & NDCG@10 
        & Recall@10 & MRR@10 & NDCG@10 \\
        \midrule
        GRU4Rec 
        & 0.6954 & 0.4055 & 0.4748 
        & 0.3851 & 0.1891 & 0.2351 
        & 0.2530 & 0.1080 & 0.1440 
        & 0.6028 & 0.2929 & 0.3660 \\
        
        BERT4Rec 
        & 0.7119 & 0.4041 & 0.4776 
        & 0.3478 & 0.1584 & 0.2027 
        & 0.2300 & 0.0930 & 0.1130 
        & 0.5490 & 0.2541 & 0.2916 \\
        
        SASRec 
        & 0.7205 & 0.4251 & 0.4958 
        & 0.4332 & 0.2325 & 0.2798 
        & 0.2710 & 0.1250 & 0.1600 
        & 0.6459 & 0.3404 & 0.4128 \\
        
        LinRec 
        & 0.7184 & 0.4316 & 0.5002 
        & 0.4270 & 0.2314 & 0.2775 
        & 0.2690 & 0.1245 & 0.1595 
        & 0.6384 & 0.3355 & 0.4073 \\
        
        LightSANs 
        & 0.7195 & 0.4314 & 0.5003 
        & 0.4406 & 0.2358 & 0.2840 
        & 0.2720 & 0.1255 & 0.1610 
        & 0.6488 & 0.3415 & 0.4142 \\
        
        SMLP4Rec 
        & 0.6753 & 0.3870 & 0.4558 
        & 0.4457 & 0.2408 & 0.2891 
        & 0.2845 & 0.1310 & 0.1670 
        & 0.6480 & 0.3484 & 0.4195 \\
        
        Mamba4Rec 
        & 0.7238 & 0.4368 & 0.5054 
        & 0.4233 & 0.2213 & 0.2689 
        & 0.2535 & 0.1085 & 0.1423 
        & 0.6488 & 0.3389 & 0.4123 \\

        ECHO 
        & 0.7215 & 0.4406 & 0.4993 
        & 0.4470 & 0.2380 & 0.2870 
        & 0.2840 & 0.1305 & 0.1665 
        & 0.6470 & 0.3460 & 0.4170 \\
        
        SIGMA 
        & 0.7235 & 0.4425 & 0.5020
        & 0.4490 & 0.2443 & 0.2915
        & 0.2890 & 0.1325 & 0.1690
        & 0.6525 & 0.3495 & 0.4215 \\

        \textbf{STAR-Rec(Ours)} 
        & \textbf{0.7260$^{\ast}$} & \textbf{0.4445}$^{\ast}$ & \textbf{0.5060}$^{\ast}$ 
        & \textbf{0.4560}$^{\ast}$ & \textbf{0.2484}$^{\ast}$ & \textbf{0.2960}$^{\ast}$ 
        & \textbf{0.2952}$^{\ast}$ & \textbf{0.1352}$^{\ast}$ & \textbf{0.1726}$^{\ast}$ 
        & \textbf{0.6609}$^{\ast}$ & \textbf{0.3537}$^{\ast}$ & \textbf{0.4266}$^{\ast}$ \\
        
        \midrule
        Improv. 
        & 0.35\% & 0.45\% & 0.80\% 
        & 1.56\% & 1.68\% & 1.54\%
        & 2.15\% & 2.04\% & 2.13\%
        & 1.29\% & 1.20\% & 1.21\% \\
        \bottomrule
    \end{tabular}}
\end{table*}
\subsection{Implementation Details}

For model optimization and training stability, we employ Adam optimizer~\cite{Adam} with a learning rate of 0.001, setting both training and evaluation batch sizes to 2048 to balance computational efficiency and model performance. Based on the characteristics and data distributions of different datasets, we adopt varying hidden dimensions: 128 for ML-1M and 64 for Amazon Beauty, Baby, and Video Games. The model architecture consists of 4 attention heads in the PMA layer, a kernel size of 4 in the SSM, and 8 MOE experts. For the Mamba component, we use 1 Mamba layer with a state dimension of 32 and an expansion factor of 2. Considering the sequence length distribution and interaction patterns, we set the maximum sequence length to 200 for ML-1M and 50 for the Amazon datasets. To mitigate overfitting and enhance model generalization, we implement dropout mechanisms with rates of 0.5 for the sparser Amazon datasets and 0.2 for ML-1M.
Following standard practices in the sequential recommendation, we train all models until convergence and evaluate them on the validation set after each epoch. We implement our model using PyTorch 2.1.1, Python 3.9, and RecBole 1.2.0~\cite{recbole[1.0],recbole[1.2.0]}. Other hyperparameters remain consistent with previous sequential recommendation works~\cite{smlp4rec,Linrec}. To ensure statistical significance, we conducted each experiment with fifty different random seeds, and all experiments were conducted on a computer with four NVIDIA 4090D GPUs with 24GB memory.

\subsection{Baselines}
We compare STAR-Rec with several state-of-the-art sequential recommendation models: (1) \textbf{GRU4Rec}\cite{GRU4Rec} leverages GRU networks to model sequential dependencies in user interaction sequences; (2) \textbf{BERT4Rec}\cite{bert4rec} adapts bidirectional transformers to better understand contextual information in user behaviors; (3) \textbf{SASRec}\cite{Kang01} utilizes a self-attention mechanism to model both long-term and short-term user preferences for accurate next-item prediction; (4) \textbf{LinRec}\cite{Linrec} modifies the dot-product operation in conventional self-attention to achieve linear complexity while maintaining high effectiveness; (5) \textbf{LightSANs}\cite{lightsan} introduces a lightweight self-attention architecture to efficiently capture sequential patterns in user behavior; (6) \textbf{SMLP4Rec}\cite{smlp4rec} adopts a simplified MLP architecture for sequential recommendation while maintaining competitive performance; (7) \textbf{Mamba4Rec}\cite{mamba4rec}, denoted as Mamba, explores the selective state space model for sequential recommendation to achieve linear computational complexity; (8) \textbf{ECHOMamba4Rec}\cite{wang2024echomamba4rec}, denoted as ECHO, enhances Mamba4Rec by introducing a bi-directional frequency-domain filter to better capture sequential patterns; (9)
\textbf{SIGMA} \cite{liu2024bidirectional} introduces a selective gated mamba architecture with a partially flipped mechanism and feature extract GRU for enhanced contextual and comprehensive short-term preference modeling in the sequential recommendation.

\subsection{Overall Performance Comparison (RQ1)}
In this subsection, we compare the performance of STAR-Rec with both traditional recommendation frameworks and state-of-the-art efficient models. The results, as shown in Table~\ref{tab:performance}, demonstrate the effectiveness of STAR-Rec on metrics Recall@10, MRR@10, and NDCG@10 in bold. According to the above table, STAR-Rec consistently outperforms all baselines across different datasets, improving the performance by 0.35$\%$-2.15$\%$. Based on these results, we conduct a comprehensive analysis of model performance:

\noindent\textbf{(1)~Traditional Models' Limitations} Traditional RNN-based models (e.g., GRU4Rec) suffer from gradient vanishing problems in capturing long-term dependencies, particularly evident when handling long sequences in the ML-1M dataset. Although Transformer-based models (e.g., BERT4Rec and SASRec) demonstrate competitive performance, they face efficiency challenges due to their quadratic computational complexity. Recent efficient models (including LinRec and LightSANs) attempt to reduce computational costs but often sacrifice model expressiveness.

\noindent\textbf{(2)~Recent Architectures' Trade-offs} Modern architectures like SMLP4Rec and Mamba4Rec show promising results but have their own strengths and limitations in different scenarios. SMLP4Rec performs poorly on long sequences, while Mamba4Rec shows the opposite characteristics: performing well on long sequences (ML-1M) but underperforming on datasets with more short sequences (Beauty and Video Games). This indicates their inability to effectively handle sequences of different lengths. Recent models like ECHO and SIGMA introduce sophisticated frequency domain filtering and partially flipped structures with gate mechanisms to address Mamba4Rec's limitations, showing promising empirical results, but their complex design may limit practical flexibility.

\noindent\textbf{(3)~STAR-Rec's Advantages} In contrast, STAR-Rec models based on the hybrid nature of long and short sequences and multi-interest characteristics in sequential recommendation, combining PMA mechanism and state space modeling to effectively capture both static item relationships and temporal dynamics, while its mixture-of-experts architecture enables specialized handling of different pattern types. This comprehensive design allows our model to adaptively handle sequential recommendation tasks across various scenarios, demonstrating superior performance.

\noindent\textbf{(4)~Performance Analysis} Notably, STAR-Rec shows larger improvements on Amazon datasets (1.20$\%$-2.15$\%$) compared to ML-1M (0.35$\%$-0.80$\%$). We hypothesize that this could be because existing models may not adequately consider the characteristics prevalent in e-commerce scenarios, such as abundant short-sequence interactions, rapidly changing user interests, and sparse interaction patterns, which are common in online shopping behaviors.

In summary, our comprehensive experiments demonstrate that STAR-Rec not only achieves state-of-the-art performance but also exhibits strong adaptability across different scenarios. The model's success can be attributed to its innovative design that effectively combines PMA with state-space modeling.

\subsection{Limited Historical Information (RQ2)}
\begin{table}[t]
    \caption{Performance comparison with different maximum sequence lengths. The best results are in \textbf{bold} with marker $\ast$ for statistical significance ($p<0.05$).}
    \label{tab:max_seq_length}
    \renewcommand{\arraystretch}{1}
    \resizebox{0.9\linewidth}{!}{
    \begin{tabular}{ccccc}
        \toprule
        Max Length & Methods & Recall@10 & MRR@10 & NDCG@10\\
        \midrule
        \multirow{5}{*}{5} & SASRec &{0.4185} &{0.2223} &{0.2685}\\\
        & LinRec & 0.3952 & 0.2102 & 0.2543\\
        & Mamba4Rec & 0.4062 & 0.1941  & 0.2440 \\
        & SIGMA & 0.4255 & 0.2315 & 0.2780\\
        & STAR-Rec &  \textbf{0.4470}$^*$ & \textbf{0.2424}$^*$ & \textbf{0.2907}$^*$ \\
        \midrule
        \multirow{5}{*}{10} & SASRec & {0.4192} &{0.2235} &{0.2694}\\\
        & LinRec & 0.4105 & 0.2186 & 0.2635\\
        & Mamba4Rec & 0.4209   & 0.2104  & 0.2622\\
        & SIGMA & 0.4395 & 0.2380 & 0.2850\\
        & STAR-Rec & \textbf{0.4515}$^*$ & \textbf{0.2487}$^*$ &  \textbf{0.2966}$^*$\\
        \midrule
        \multirow{5}{*}{20} & SASRec  & 0.4183 & 0.2246 & 0.2703  \\
        & LinRec & 0.4175 & 0.2238 & 0.2698\\
        & Mamba4Rec&  0.4251 & 0.2168 & 0.2658 \\
        & SIGMA & 0.4434 & 0.2412 & 0.2885\\
        & STAR-Rec &  \textbf{0.4504}$^*$ & \textbf{0.2456}$^*$ & \textbf{0.2940}$^*$\\
        \bottomrule
    \end{tabular}}
\end{table}
In this subsection, we investigate model robustness by evaluating performance under extreme sequence length settings. Varying the maximum allowable sequence length offers insights into how models handle different levels of historical interaction information.

Specifically, we truncate or filter user interaction sequences by setting maximum lengths to 5, 10, and 20, where sequences exceeding these limits are truncated from the most recent interactions. This investigation is particularly meaningful as it represents more challenging scenarios where models must make recommendations with significantly limited historical information.

The experimental results in Table~\ref{tab:max_seq_length} decisively demonstrate STAR-Rec's overwhelming superiority. When drastically reducing the maximum length to just five interactions (compared to 100 in main experiments), where models must make predictions with extremely limited historical context, STAR-Rec still achieves remarkable performance with Recall@10 of 0.4470, significantly outperforming both traditional SASRec (0.4185) and recent SIGMA (0.4255). This outstanding performance under severe information constraints highlights STAR-Rec's powerful modeling capability.
While SIGMA and Mamba4Rec improve their performance as we increase the maximum length to 10 and 20, STAR-Rec maintains its clear advantage across all settings, demonstrating its exceptional robustness and adaptability to information-constrained scenarios.

\subsection{Ablation Study (RQ3)}
\begin{figure}[t]
    \centering
    \includegraphics[width=0.95\linewidth]{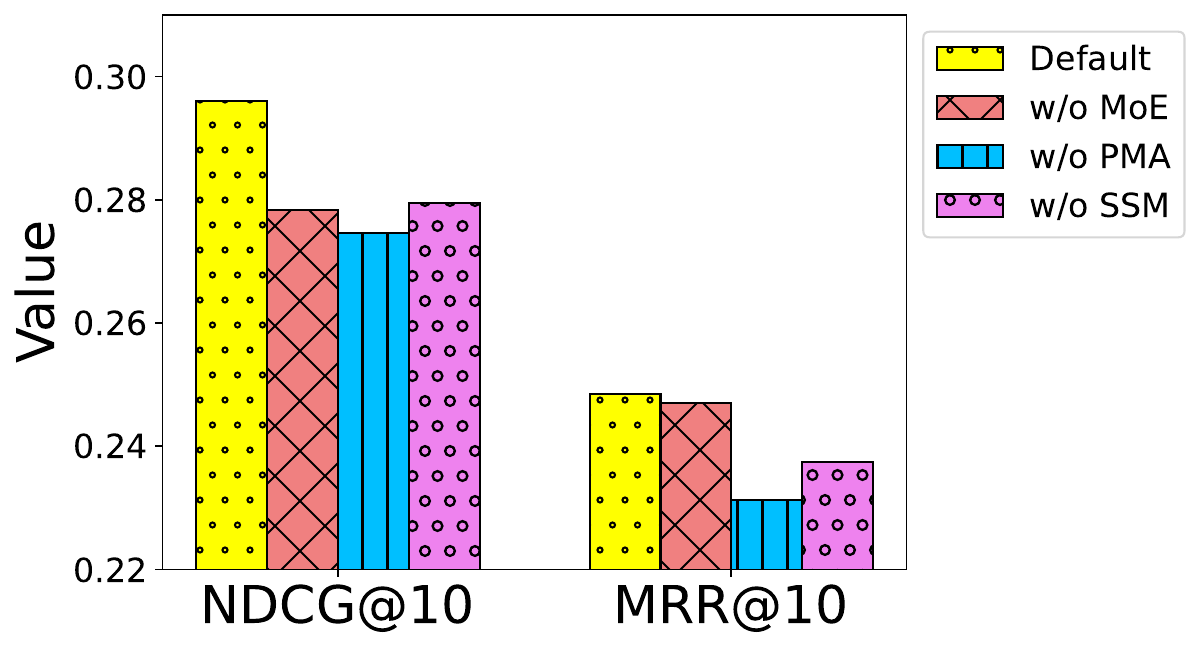}
    \caption{Ablation study of different components in our model. We evaluate the effectiveness of MoE, PMA, and SSM by removing them, respectively.}
    \vspace{-5pt}
    \label{fig:ablation}
    \vspace{-5pt}
\end{figure}

\begin{table}[t]
\caption{Parameter study of different layer numbers on Beauty dataset. The single-layer model can achieve the best performance while maintaining lower computational costs.}
\centering
\resizebox{0.85\columnwidth}{!}{
\begin{tabular}{c|ccc}
\hline
\#layers & Recall@10 & MRR@10 & NDCG@10 \\
\hline
1  & 0.4560 & 0.2484 & 0.2960 \\
2 & 0.4525 & 0.2465 & 0.2955 \\
3  & 0.4412 & 0.2401 & 0.2872 \\
4 & 0.4325 & 0.2359 & 0.2814 \\
\hline
\end{tabular}
}
    \vspace{-8pt}
\label{tab:layer_study}
\end{table}

In this subsection, we conduct comprehensive ablation experiments to evaluate the contribution of each key component in STAR-Rec. Following our model design philosophy of integrating state-space modeling with similarity-enhanced attention, we examine three critical modules: MoE, PMA, and SSM. We create three variants by removing each component individually and report their performance on the Beauty dataset in Figure~\ref{fig:ablation}.

The experimental results provide strong evidence for the effectiveness of our design choices. First, removing the MoE prediction layer leads to a noticeable performance drop (e.g., NDCG@10 decreases from 0.2960 to 0.2783), indicating its importance in handling diverse behavioral patterns captured by both similarity attention and temporal modeling. The PMA mechanism proves to be particularly crucial, as its removal results in the most significant performance degradation across all metrics (MRR@10 drops from 0.2484 to 0.2313). 
This substantial impact highlights the vital role of explicitly modeling item relationships and static user preferences, which cannot be effectively captured by temporal dynamics alone. Finally, the SSM component also demonstrates its value in modeling temporal evolution, with its removal causing a substantial performance decrease (NDCG@10 drops from 0.2960 to 0.2795).
These ablation results clearly validate our architectural design of combining similarity-based preference modeling with temporal dynamics through state-space models. 

\subsection{Parameter Sensitivity Analysis (RQ4)}
\begin{figure}[t]
\centering
\includegraphics[width=0.85\linewidth]{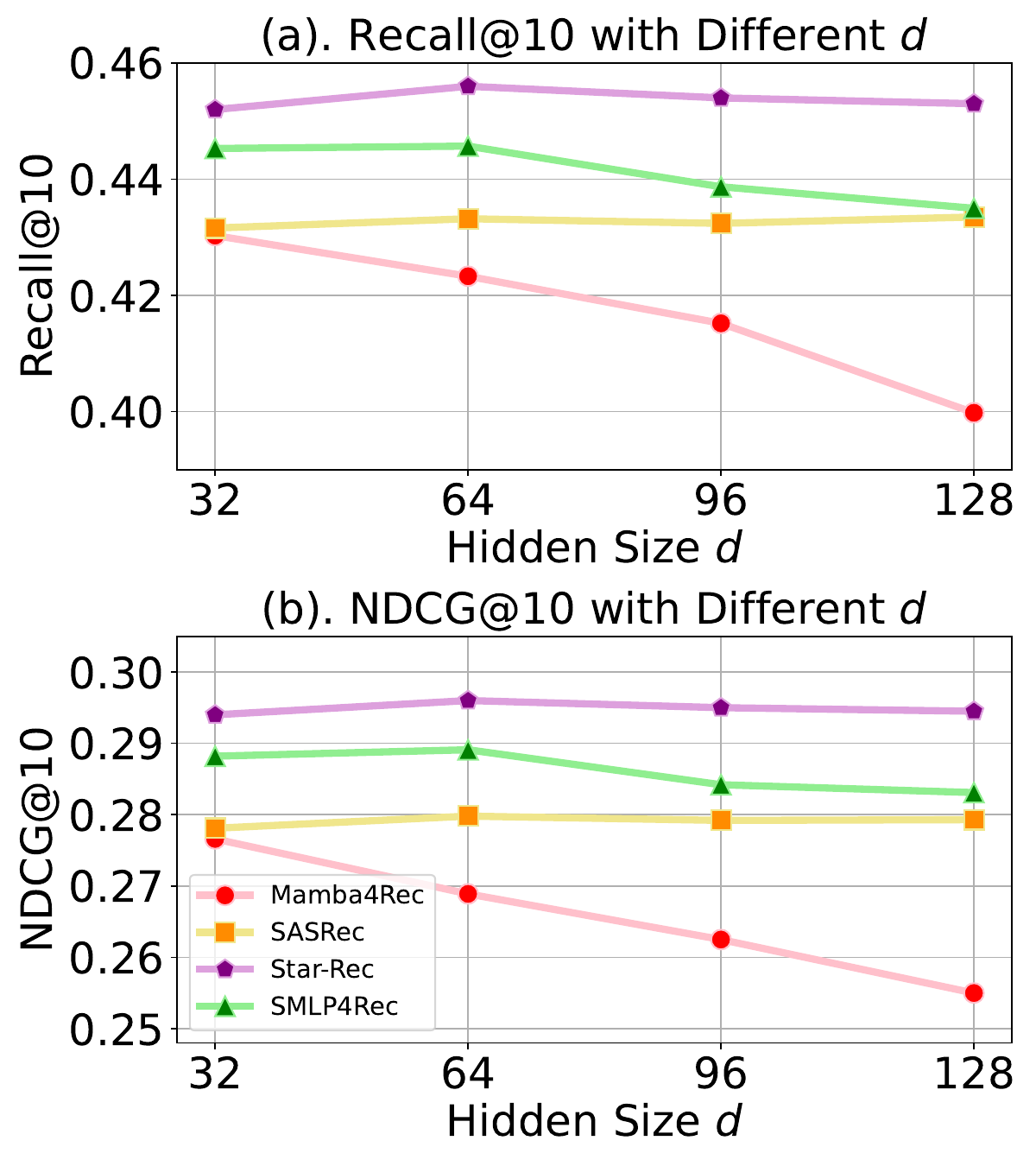}
\vspace{-15pt}
\caption{Impact of hidden size $d$ on model performance.}
\label{fig:parameter_d}
   \vspace{-5mm}
\end{figure}

\begin{table}[!t]
    \caption{Efficiency comparison between different sequential recommendation methods. Results show inference time per mini-batch and training time per epoch.}
    \vspace{-5pt}
    \label{tab:efficiency}
    \renewcommand{\arraystretch}{1}
    \resizebox{0.85\linewidth}{!}{
    \begin{tabular}{cccc}
        \toprule
        Datasets & Model & Infer. & Training\\
        \multirow{9}{*}{Beauty} & BERT4Rec & 1368ms & 12.8s/epoch\\
        \midrule
            & SASRec & 446ms & 8.3s/epoch\\
            & LinRec & 337ms & 5.7s/epoch\\
            & LightSANs & 425ms & 7.1s/epoch\\
            & SMLP4Rec & 358ms & 5.4s/epoch\\
            & Mamba4Rec & 353ms & 4.7s/epoch\\
            & ECHO & 403ms & 5.3s/epoch\\
            & SIGMA & 375ms & 4.9s/epoch\\
            & STAR-Rec(Ours) & 368ms & 4.2s/epoch\\
            \midrule
        \multirow{9}{*}{Video Games} & BERT4Rec & 1290ms & 15s/epoch\\
            & SASRec & 406ms & 9.6s/epoch\\
            & LinRec & 327ms & 6.6s/epoch\\
            & LightSANs & 369ms & 8.3s/epoch\\
            & SMLP4Rec & 389ms & 9.1s/epoch\\
            & Mamba4Rec & 309ms & 5.6s/epoch\\
            & ECHO & 385ms & 7.8s/epoch\\
            & SIGMA & 295ms & 5.2s/epoch\\
            & STAR-Rec & {297ms} & {5.7s/epoch}\\
        \bottomrule
    \end{tabular}
    }
    \vspace{-10pt}
\end{table}

From Table \ref{tab:layer_study}, we investigate the impact of model depth on Star-Rec through extensive experiments. Table~\ref{tab:layer_study} reveals that a single-layer Star-Rec achieves comparable performance to its two-layer variant, albeit the latter consumes more computational resources. More importantly, we observe significant performance degradation when increasing the model depth to three or four layers. This phenomenon can be attributed to several factors: the single-layer architecture already effectively captures temporal dynamics in sequences, while adding more layers not only introduces excessive parameters and complexity but also leads to optimization challenges and gradient vanishing issues. These findings demonstrate that deeper architectures do not necessarily yield better performance in sequential recommendation tasks. Instead, the key lies in designing an appropriate single-layer structure for effective temporal information encoding while balancing model expressiveness and efficiency.

Through Figure \ref{fig:parameter_d}, we examine the impact of hidden dimensions on model effectiveness. The experimental results show that Star-Rec consistently outperforms the baselines across different hidden sizes, particularly achieving substantial gains at $d=64$ compared to Mamba4Rec's state-space modeling and SASRec's attention mechanism. Notably, both Mamba4Rec and SMLP4Rec show sensitivity to the hidden dimension, with their performance slightly declining as the dimension increases. This suggests these models are more sensitive to the choice of hidden dimension, potentially due to their structural characteristics in processing sequential information.

\subsection{Efficiency Analysis (RQ5)}


In this subsection, we analyze STAR-Rec's efficiency through inference time, training time, and GPU memory usage. As shown in Table~\ref{tab:efficiency}, traditional Transformer-based models (BERT4Rec, SASRec) consume substantial computational resources, while recent Mamba-based models demonstrate improved efficiency.
STAR-Rec maintains efficiency metrics comparable to state-of-the-art models across all datasets. On all tested datasets, including Beauty and Video Games, our model demonstrates comparable inference time and memory consumption with current leading methods such as Mamba4Rec. These results indicate that STAR-Rec achieves competitive efficiency while delivering superior recommendation performance, as shown in RQ1.
\section{Related Works}

\noindent\textbf{Transformers and RNNs for Sequential Recommendation}
Sequential recommendation has evolved significantly from traditional methods to deep learning-based solutions~\cite{Frequency23,DL4,Xavier,sse-pt,zhao2023embedding,FMLP,strec,MLM4Rec,PEPNet,mb-str,lightsan,autoseqrec,HRNN,zhao2023user}. Early approaches like TransRec~\cite{DMAN} and matrix factorization methods~\cite{koren2009matrix} focused on modeling user-item interactions through conventional data mining techniques, but they struggled with capturing multiple user behaviors and faced efficiency challenges with longer sequences. This led to the emergence of deep learning methods, particularly Transformers and RNNs. Transformer-based models like SASRec~\cite{Kang01} leveraged multi-head attention mechanisms for sequence modeling, while BERT4Rec~\cite{bert4rec} employed bidirectional transformers to capture contextual information. LinRec~\cite{Linrec} further improved efficiency by introducing linear complexity attention mechanisms. Despite their effectiveness, these transformer-based models suffer from quadratic computational complexity when modeling long sequences. RNN-based approaches like GRU4Rec~\cite{GRU4Rec} provided linear computational complexity but showed limited effectiveness in sequential recommendations. To address this, STAR-Rec combines preference-aware attention and state-space modeling to handle variable-length sequences while maintaining efficiency.

\noindent\textbf{State Space Models for Sequential Recommendation}
Recently, state-space-models (SSMs) have demonstrated remarkable effectiveness in sequence modeling tasks due to their superior capability in capturing temporal dynamics and hidden patterns~\cite{GLINTours25,HiPPOs21,16Dual,gu2023mamba,qu2024survey,dao2024transformers,MambaRec,wang2024echomamba4rec,cao2024mamba4kt,liu2024bidirectional,yang2024uncovering,Visionzhu}. Mamba4Rec~\cite{mamba4rec} pioneered this direction by demonstrating improved efficiency while maintaining competitive performance through its selective state space modeling. Following this, ECHO-Mamba4Rec~\cite{wang2024echomamba4rec} advanced the field by combining bidirectional Mamba with frequency-domain filtering for more accurate pattern capture. RecMamba~\cite{yang2024uncovering} demonstrated Mamba's capability in handling lifelong scenarios, while Mamba4KT~\cite{cao2024mamba4kt} adapted the architecture for knowledge tracing applications. Most recently, SIGMA~\cite{liu2024bidirectional} attempted to address Mamba's limitations in context modeling and short sequence handling through a bi-directional structure with selective gating mechanisms. These approaches face challenges in balancing long/short-term sequence modeling and pattern diversity in recommendation scenarios. STAR-Rec addresses this through preference-aware attention and state-space modeling via sequence-level mixture-of-experts, effectively handling diverse item relationships and varying sequence patterns.

\vspace{-9pt}
\section{Conclusion}
In this paper, we propose STAR-Rec, a novel sequential recommendation framework that effectively integrates preference-aware attention with state-space modeling through a sequence-level mixture-of-experts framework. Our work addresses three key characteristics in the sequential recommendation by: (1) employing preference-aware attention to capture item relationships, (2) utilizing state-space modeling for efficient sequence processing, and (3) incorporating a mixture-of-experts architecture that adaptively handles different behavioral patterns. Extensive experiments demonstrate that STAR-Rec consistently outperforms state-of-the-art methods, with ablation studies validating each component's contribution. Future work could explore handling more complex sequential patterns and incorporating additional contextual information.

\section*{Acknowledgement}
This research was partially supported by Research Impact Fund (No.R1015-23), Collaborative Research Fund (No.C1043-24GF), Huawei (Huawei Innovation Research Program, Huawei Fellowship), Tencent (CCF-Tencent Open Fund, Tencent Rhino-Bird Focused Research Program), Alibaba (CCF-Alimama Tech Kangaroo Fund No. 2024002), Ant Group (CCF-Ant Research Fund), and Kuaishou.
\bibliographystyle{ACM-Reference-Format}
\bibliography{sample-base}

\end{document}